\documentclass{article}
\usepackage{graphicx}  
\usepackage{amsmath}   
\usepackage[compress]{cite}
\usepackage{amssymb}   
\usepackage{bm} 
\usepackage{dcolumn}
\usepackage{color}
\usepackage{mathrsfs}
\usepackage{amsfonts}
\usepackage{varioref}
\usepackage{textcomp}

\RequirePackage[colorlinks,citecolor=blue,urlcolor=magenta,linkcolor=blue]{hyperref}
\allowdisplaybreaks
\labelformat{section}{Section #1} 
\labelformat{subsection}{Section #1} 
\labelformat{subsubsection}{Section #1}
\labelformat{subsubsubsection}{Section #1}
\labelformat{equation}{Eq.~(#1)} 
\labelformat{figure}{Fig.~#1} 
\labelformat{subfigure}{Fig.~\thefigure#1} 
\labelformat{table}{Table~#1} 
\labelformat{appendix}{Appendix #1}
\newcounter{mnotecount}[section]

\renewcommand{\themnotecount}{\thesection.\arabic{mnotecount}}

\newcommand{\mnotex}[1]
{\protect{\stepcounter{mnotecount}}$^{\mbox{\footnotesize
$
\bullet$\themnotecount}}$ \marginpar{
\raggedright\small\em
$\!\!\!\!\!\!\,\bullet$\themnotecount: #1} }

\begin{document}
\title{Galactic Pure Lovelock Blackholes: Geometry, stability, and Hawking temperature }
\author{ Chiranjeeb Singha\footnote{chiranjeeb.singha@saha.ac.in}$~^{1}$ and Shauvik Biswas
\footnote{shauvikbiswas2014@gmail.com}$~^{2}$ \\
$^{1}$\small{Theory Division, Saha Institute of Nuclear Physics, Kolkata 700064, India}\\ 
$^{2}$\small{School of Physical Sciences, Indian Association for the Cultivation of Science, Kolkata-700032, India}\\
}
\date{\today}
\maketitle

\begin{abstract}
In this article, we will first-time model galactic black holes in pure Lovelock gravity. Even though
working with higher spacetime dimensions, we assume (implicitly) the Hernquist-type mass profile for
the galaxy in such a way that the horizon structure of a pure LoveLock black hole remains intact. In this way, we will model the galactic pure Lovelock black hole with arbitrary dimension ($d$) and order ($N$). Then, we will specialize this technique for critical dimension $d=3N+1$. We want to see how the galactic parameters affect the time domain single, quasinormal modes, photon sphere, innermost stable circular orbits (ISCO), and shadow radius. The time domain signal may allow us to identify the galactic parameters as well as to distinguish them from their isolated pure Lovelock counterparts if it is observed in future generations of gravitational wave measurements. We also calculate Hawking temperature for the same setup and want to see how Hawking's temperature will be affected due to the presence of a galaxy. It shows that the presence of a galactic halo can quench Hawking temperature.
\end{abstract}
\section{Introduction} 
 Numerous experiments, including investigations into the flatness of galaxies' rotation curves \cite{Rubin1970RotationOT, cowie1986virial, Borriello:2000rv, Persic:1995ru}, the dynamics of hot gas in clusters \cite{Briel:1997hz}, and the phenomenon of gravitational lensing \cite{SDSS:2005sxd}, among others, reveal that 95 percent mass of a galaxy originates from non-baryonic matter, commonly known as dark matter \cite{Clowe:2006eq, Freese:2008cz}. Thus, in our universe, isolated objects do not exist. Any compact object, whether a black hole or an exotic compact object (ECO), must coexist with dark matter, influencing spacetime geometry.

A recent comprehensive study \cite{Cardoso:2021wlq} presented a fully relativistic analysis, deriving the spherically symmetric metric of a black hole spacetime within the presence of a galactic matter distribution following the Hernquist-type density profile \cite{hernquist1990analytical}:
\begin{equation}\label{Hernquist-density}
\rho(r)= \frac{M a}{2 \pi r (r+a)^3}~.
\end{equation}
Here, $M$ represents the mass of the dark matter halo, and $a$ is a characteristic length scale associated with the dark matter distribution within the galaxy. This density profile is a motivation for determining the mass profile of a galactic black hole, which can be described as follows: 
\begin{align}\label{Cardoso-Result-1}
m(r)=M_{\textrm{BH}}+\frac{M r^{2}}{(r+a)^{2}}\left(1-\frac{2M_{\textrm{BH}}}{r}\right)^{2}~, 
\end{align}
where $M_{\textrm{BH}}$ is mass of the central black hole. The resulting geometry from the mass above profile bears a resemblance to the Einstein cluster \cite{7bb06a79-8225-31c6-88c3-0c4f8a76b072}. Importantly, this mass profile preserves the black hole horizon's existence, even in a galaxy's presence \footnote{To explore galactic black hole solutions featuring different mass profiles, refer to \cite{Konoplya:2022hbl}.}. It is worth noting that the galactic matter also provides an effective shielding mechanism when the central black hole possesses an electric charge \cite{Feng:2022evy}. Building on the insights gained from the fully relativistic analysis of galactic black holes, in this paper, we focus on exploring the scenario of a pure Lovelock black hole in the center of a galaxy. Pure Lovelock gravity is a theory characterized by a modified kinetic term in dimensions $d > 4$, formulated through a polynomial in the Riemann curvature of order N. When N equals 1, the theory simplifies to the linear term, representing scalar curvature and coinciding with Einstein's General Relativity (GR). For $d = 4$, the only viable option is N = 1. Generally, d and N are independent, except for the condition that $N > 1$ is permissible in dimensions $d > 4$, and it consistently holds that $N < [(d-1)/2]$. Even though higher curvature terms exist in the Lagrangian for pure Lovelock theories, the field equations remain second-order \cite{Gannouji:2019gnb, Dadhich:2015nua, Dadhich:2012cv, Gannouji:2013eka, Dadhich:2015ivt, Padmanabhan:2013xyr, Dadhich:2015lra}. Rather than engaging with elevated spacetime dimensions, we implicitly adopt the Hernquist-type mass profile for the galaxy. This approach ensures that the fundamental horizon structure of a pure Lovelock black hole remains undisturbed. We want to see how the galactic parameters affect the time domain single, quasinormal modes, photon sphere, ISCO, and shadow radius. The time domain single may allow us to identify the galactic parameters as well as to distinguish them from their isolated pure Lovelock counterparts and the GR counterparts theoretically or experimental
observation but the effect of extra dimension in pure Lovelock gravity must be reduced to $d = 4$ by some mechanism (e.g., compactification). We also calculate Hawking temperature and want to see how Hawking's temperature will be affected due to the presence of a galaxy. 

This paper is organized as follows: First, in \ref{Einstein}, we briefly review the geometry of general relativistic black holes in galactic centers. Next, in \ref{lovelock}, we present the construction of a pure Lovelock black hole with arbitrary dimension ($d$) and order ($N$) in a galactic center. Then, we specialize this technique for the critical dimension $d=3N+1$. The spacetime properties, including the photon sphere, ISCO, shadow radius, and Hawking temperature, are then calculated.
Moving on to \ref{sec-4}, we analyze the massless scalar perturbation and determine the effective potential experienced by it. We calculate the scalar quasinormal modes by employing the 6th-order Wentzel-Kramers-Brillouin (WKB) approximation \cite{Konoplya:2004ip}. Then, we solve the master equation in the time domain, enabling us to obtain the ringdown waveform. Finally, our conclusions in \ref{sec-5} encompass a comprehensive discussion of our findings and an outline of potential avenues for future research.

\emph{Notations and Conventions:} We will set $c=\hbar=G=1$ in our calculations. Additionally, our metric will follow the mostly positive signature convention, which means that in four spacetime dimensions, the Minkowski metric takes the form $\eta_{\mu \nu}=\textrm{diag.}(-1,+1,+1,+1)$. Throughout the paper, `prime' denotes the derivative with respect to $r$.
%

\section{Review Of The geometry of GR black holes in galactic centers}\label{Einstein}
First, we review the geometry of a Schwarzschild black hole. The metric for the Schwarzshild black hole is given by,
\begin{equation}
    ds^2=- f(r) dt^2+ f(r)^{-1} dr^2+r^2 d \Omega^2~,\label{Schwarzschild}  
\end{equation}
where $f(r)=\left(1-2 M_{BH}/r\right)$. Here $M_{BH}$ is the mass of the black hole. One can calculate the radius of the photon sphere, the shadow radius, and the ISCO from this metric. The calculation of $r_{ph}$ involves determining the effective potential encountered by a photon navigating through this spacetime, which is given by,
\begin{align}\label{photon-schw}
    V_{\textrm{eff}}^{\textrm{ph}}=\frac{f(r)}{r^{2}}~.
\end{align}
The photon sphere is the unstable circular null geodesics that are obtained by setting the derivative of the effective potential to zero. This yields the algebraic equation $r f^{\prime}= 2 f$ \cite{Cardoso:2008bp}, enabling the determination of the photon sphere's radius ( $r_{ph}$). For the Schwarzschild spacetime, it is given by,
\begin{equation}
   r_{\rm ph}=3 M_{BH}~.\label{Photon}  
\end{equation}
 To determine the innermost stable circular orbit we consider a motion of a massive particle in spacetime and impose the following three conditions on the effective potential experienced by the test massive particle: $V_{\textrm{eff}}=0$, $V^{\prime}_{\textrm{eff}}=0$, and $V^{\prime\prime}_{\textrm{eff}}=0$. The solution of these algebraic conditions leads to the following solution for the Schwarzschild spacetime,
\begin{align}
    r_{\textrm{ISCO}}=6M_{\textrm{BH}}~.
\end{align}
The calculation of $r_{sh}$ involves determining the critical impact parameter associated with null geodesics which is defined by the relation
\begin{align}\label{sch-sh}
    \frac{1}{r^{2}_{\textrm{sh}}}= V_{\textrm{eff}}^{\textrm{ph}}(r_{ph})~.
\end{align}
This for Schwarzschild spacetime leads to, 
\begin{align}
   r_{sh}=3\sqrt{3}M_{BH}~. 
\end{align}

Then, we consider a GR black hole in the galactic center. Furthermore, the arrangement is expected to remain static, meaning no radial outflow should be present. Thus, for constructing the metric of this system, we choose a perfect fluid of energy-momentum tensor $T^{\mu}_{\nu}=\textrm{diag.} \left(-\rho, 0, P_{T}, P_{T}\right)$. We also consider the spherical symmetric metric for this system in the following manner,

\begin{equation}
    ds^2=- f (r) dt^2+ \left(1-\frac{2 m(r)}{r}\right)^{-1} dr^2+r^2 d \Omega^2~.\label{Schwarzschild_gal}  
\end{equation}
Here we consider the mass profile as $m(r)=M_{BH}+ \frac{M r^2}{(r+a)^2}\left(1-2 M_{BH}/r\right)^{2}$ where $M$ is the total mass of the ``halo" and $a$ a typical length scale associated with the dark matter distribution in the galaxy \cite{Cardoso:2021wlq}. Significantly, this mass profile maintains the structure of the GR black hole horizon, even when the GR black hole is situated in the galactic center. Now from the radial component of Einstein equation $G^{r}{ }_{r}=8 \pi T^{r}{}_{r}$ we get,
\begin{equation}\label{presser}
    \frac{r f^{\prime}}{2 f}=\frac{m(r) \rho}{r- 2 m(r)}~.
\end{equation}
The solution of the above equation provides us with the following,
\begin{eqnarray}
    f &=&\left(1- \frac{2 M_{BH}}{r}\right) e^{\gamma}~, \label{Black_gal_f}\\
    \gamma &=& -\pi \sqrt{\frac{M}{\xi}}+2 \sqrt{\frac{M}{\xi}} \tan^{-1} \frac{r+a-M}{\sqrt{M \xi}}~,\\
    \xi &=& 2 a -M +4 M_{BH}~.
\end{eqnarray}

From the mass profile, we can derive the matter density profile. The temporal  component of the Einstein equation, given by $G^{t}{ }_{t}=8 \pi T^{t}{}_{t}$, offers us the matter density in the subsequent manner \cite{Cardoso:2021wlq},
\begin{eqnarray}
    4 \pi \rho &=& \frac{m'(r)}{r}\nonumber\\
    \Rightarrow \rho &=&\frac{2 M (a+2 M_{BH})(1-2M_{BH}/r)}{4 \pi r (a+r)^3}~.
\end{eqnarray}
The tangential pressure can be obtained from the Bianchi identities and is characterized by its expression as stated in \cite{Cardoso:2021wlq},
\begin{eqnarray}
    2 P_{T} = \frac{m(r)\rho}{r- 2 m(r)}~.
\end{eqnarray}
 From the metric (\ref{Schwarzschild_gal}), one can calculate the radius of the photon sphere, the shadow radius, and the ISCO. Upto the leading order in $\left(\frac{M_{BH}}{a}\right)$, the radius of the photon sphere is given by \cite{Cardoso:2021wlq},
\begin{equation}
   r_{\rm ph}=3 M_{BH}\bigg(1+\frac{M M_{BH}}{a^2}\bigg)~.\label{Photon_gal}  
\end{equation}
Note that in the limit of $ M\to 0$, we get back the result of the photon sphere of the isolated Schwarzschild black hole. Furthermore,  the presence of a dark matter halo increases the radial location of the photon sphere, compared to the case of an isolated Schwarzschild black hole. The radial coordinate solution of ISCO is given by \cite{Cardoso:2021wlq},
\begin{align}\label{Schwarzschildgal-ISCO}
r_{\rm ISCO}=6M_{BH}\bigg(1-\frac{32 M M_{BH}}{a^2}\bigg)~.
\end{align}
Here also, as $M$ approaches zero, we recover the ISCO outcome for the isolated Schwarzschild black hole. Additionally, introducing a dark matter halo reduces the radial location of the ISCO position relative to the isolated Schwarzschild black hole scenario. The shadow radius associated with the galactic blackhole is given by \cite{Cardoso:2021wlq},
\begin{align}\label{Schwarzschildgal-Shadow}
r_{\rm sh}=3\sqrt{3}M_{BH}\bigg[1+\frac{M}{a}+\frac{M(5M-18 M_{BH})}{6 a^2}\bigg]~.
\end{align}
Similarly, in the limit of $ M\to 0$, we get back the result of the shadow radius of the isolated Schwarzschild black hole. Also, in the presence of a dark matter halo, the shadow radius will be larger than that of an isolated Schwarzschild black hole.

\section{The geometry of pure Lovelock black holes in galactic centers.}\label{lovelock}

In this section, we first briefly review the geometry of a pure Lovelock black hole of Lovelock order N in d dimensional spacetime.  The metric for a pure Lovelock black hole of Lovelock order N in d dimensional spacetime is given by \cite{Gannouji:2019gnb, Dadhich:2015nua, Dadhich:2012cv, Gannouji:2013eka, Dadhich:2015ivt, Padmanabhan:2013xyr, Dadhich:2015lra},
\begin{equation}\label{Lovelock}
    ds^2= -\left(1-\left(\frac{2^{N} M_{BH}}{r^{d-2 N -1}}\right)^{1/N}\right) dt^2+\left(1-\left(\frac{2^{N} M_{BH}}{r^{d-2 N -1}}\right)^{1/N}\right)^{-1} dr ^2 + r^2 d\Omega^2_{d-2} ~.
\end{equation}
The ADM mass is represented as $M_{BH}=(M_{BH}^{1/N})^N$ in this context, and the given metric corresponds to a solution of vacuum Einstein's equation.
Then we present a modeling approach for the d-dimensional galactic pure Lovelock black hole with a Lovelock order N. The arrangement is expected
to remain static, \emph{i.e.,} no radial outflow should be present. Thus, to construct the metric for this system, we adopt a perfect fluid with an energy-momentum tensor represented as $T^{\mu}_{\nu}=\textrm{diag.} \left(-\rho, 0, P_{T}, P_{T}\right)$. The metric describing the galactic pure Lovelock black hole is assumed to possess spherical symmetry and is formulated as follows:
\begin{equation}
    ds^2=- f (r) dt^2+ \left(1-\left(\frac{2^{N} m(r)}{r^{d-2 N -1}}\right)^{1/N}\right)^{-1} dr^2+r^2 d \Omega{}^2_{d-2}~.\label{generalLovelock_gal}  
\end{equation}
Here $f(r)$ and $m(r)$ are the two unknown quantities. Motivated by the galactic black hole, here we choose $f(r)$ in the following way,
\begin{eqnarray}
    f &=&\left(1-\left(\frac{2^{N} m(r)}{r^{d-2 N -1}}\right)^{1/N}\right) e^{\gamma}~, \label{general_lovelock_gal}\nonumber\\
    \gamma &=& -\pi \sqrt{\frac{M}{\xi}}+2 \sqrt{\frac{M}{\xi}} \tan^{-1} \frac{r+a-M}{\sqrt{M \xi}}~,\nonumber\\
    \xi &=& 2 a -M +4 M_{BH}~.\label{Lovelock_gal_f}
\end{eqnarray}
Note that, unlike the previous section, here we have fixed $g_{tt}$ by implicitly assuming the Harnquint type mass profile through the redshift factor, that is, via $e^{\gamma}$. That is, even though
working with higher spacetime dimensions, we assume (implicitly) the Hernquist-type mass profile for
the galaxy in such a way that the horizon structure of a pure LoveLock black hole remains intact. To be consistent, one may call this choice as ``Harnquist-type redshift".

 We get the following equation from the radial component of the Einstein equation $G^{r}{ }_{r}=8 \pi T^{r}{}_{r}$,
\begin{eqnarray}
    \frac{r f^{\prime}}{2 f}=\frac{(d-2 N-1)}{N}\frac{(m(r))^{1/N}}{r^{d-2 N-1}- 2 (m(r))^{1/N}}~.\label{pressure}
\end{eqnarray}
By utilizing the above equations (\ref{Lovelock_gal_f} and \ref{pressure})), we can determine the mass profile. To mimic galactic observations, we make the assumption $a>M\gg M_{BH}$\cite{Navarro:1995iw}. By solving the equations above (\ref{Lovelock_gal_f} and \ref{pressure}) and considering terms up to $\mathcal{O}(1/a^{2})$, we obtain the mass profile for the $d=3N+1$ dimensional galactic pure Lovelock black hole with Lovelock order N (see \ref{app-1} for a comprehensive mass profile for the cases of $d=7$ with $N=2$ and $d=10$ with $N=3$) as,
\begin{eqnarray}
    m(r)=M_{BH}+\frac{N M r^2 (M_{BH})^{\frac{N-1}{N}} \left(1-\frac{2 M_{BH}^{1/N}}{r}\right)^2}{a^2}~.\label{Mass_profile}
\end{eqnarray}
Importantly, this mass distribution ensures the persistence of the horizon of the pure Lovelock black hole, even when the
pure Lovelock black hole is situated in the galactic center. We first consider a seven-dimensional pure Lovelock black hole of Lovelock order two in the galactic center. Now from \ref{general_lovelock_gal}, we get the expression of $f$ for $d=7$ and $N=2$ as,
\begin{eqnarray}\label{Lovelock7_gal_f}
    f &=&\left(1- \frac{2 \sqrt{M_{BH}}}{r}\right) e^{\gamma}~,
    \end{eqnarray}
 and from \ref{Mass_profile}, we get the mass profile as,
\begin{equation}\label{massprofile_lovlock7} 
    m(r)= M_{BH}+\frac{2 M  r^2 \sqrt{M_{BH}}\bigg(1-\frac{2 \sqrt{M_{BH}}}{r}\bigg)^2}{a^2}~. 
\end{equation}
For the mass profile described above, up to $\mathcal{O}(1/a^{2})$, the $g^{rr}$ component of the metric is expressed as follows:
\begin{equation}\label{g_lovlock7}
g^{rr}\equiv g (r) =\left(1-\frac{2 \sqrt{M_{BH}}}{r}\right) \left[1-\frac{2 M r}{a^2} \left(1-\frac{2 \sqrt{M_{BH}}}{r}\right) \right] ~. 
\end{equation}
Calculating the matter density profile is achievable through the mass profile. The matter density, in a specific form, can be determined utilizing the temporal  component of the Einstein equation $G^{t}{}_{t}=8 \pi T^{t}{}_{t}$ as follows,
\begin{eqnarray}
    4 \pi \rho &=& \frac{m'(r)}{r^5}\nonumber\\
    \Rightarrow \rho &=&\frac{M \sqrt{M_{BH}}\left( r-2 \sqrt{M_{BH}}\right)}{\pi  a^2 r^5}~.
\end{eqnarray}
Deriving the tangential pressure is facilitated by the Bianchi identities \cite{Chakraborty:2021dmu}, with the expression for the tangential pressure taking the following form,
\begin{eqnarray}
    5 P_{T} = \frac{(m(r))^{1/2}\rho}{r- 2 (m(r))^{1/2}}~.
\end{eqnarray}
One can calculate the radius of the photon sphere, the shadow radius, and the ISCO for a galactic seven-dimensional pure Lovelock black hole of Lovelock order two. Upto the leading order in $\left(\frac{\sqrt{M_{BH}}}{a}\right)$, the radius of the photon sphere is given by,
\begin{equation}
   r_{\rm ph}=3 \sqrt{M_{BH}}\bigg(1+\frac{M \sqrt{M_{BH}}}{a^2}\bigg)~.\label{Lovelockgal_Photon}  
\end{equation}
Note that in the limit of $ M\to 0$, we return the result of the photon sphere of the isolated seven-dimensional pure Lovelock black hole of Lovelock order two. Also, the presence of a dark matter halo increases the radial location of the photon sphere, compared to the case of an isolated seven-dimensional pure Lovelock black hole of Lovelock order two. The radial coordinate solution of ISCO is given by,
\begin{align}\label{Lovelock-ISCO}
r_{\rm ISCO}=6\sqrt{M_{BH}}\bigg(1-\frac{32 M \sqrt{M_{BH}}}{a^2}\bigg)~.
\end{align}
Here also, as $M$ approaches zero, we recover the ISCO outcome of the isolated seven-dimensional pure Lovelock black hole of Lovelock order two. Furthermore, introducing a dark matter halo reduces the ISCO position compared to the case of an isolated seven-dimensional pure Lovelock black hole of Lovelock order two.
The shadow radius associated with the galactic seven-dimensional pure Lovelock blackhole of Lovelock order two is given by,
\begin{align}\label{Lovelock-Shadow}
r_{\rm sh}=3\sqrt{3 M_{BH}}\bigg[1+\frac{M}{a}+\frac{M(5M-18 \sqrt{M_{BH}})}{6 a^2}\bigg]~.
\end{align}
Likewise, as $M$ tends towards zero, we arrive at the outcome for the shadow radius corresponding to the isolated seven-dimensional pure Lovelock black hole of Lovelock order two. Moreover,  in the presence of a dark matter halo, we show that the shadow radius will be larger than that of an isolated seven-dimensional pure Lovelock black hole of Lovelock order two.

We then consider a ten-dimensional pure Lovelock black hole of Lovelock order three in the galactic center. Now
from \ref{general_lovelock_gal}, we get the expression of $f$ for $d=10$ and $N=3$ as,
\begin{eqnarray}\label{Lovelock10_gal_f}
    f &=&\left(1- \frac{2 M_{BH}^{1/3}}{r}\right) e^{\gamma}~,
\end{eqnarray}
and from \ref{Mass_profile}, we get the mass profile as,
\begin{equation}\label{massprofile_lovlock10}
    m(r)= M_{BH}+\frac{3 M  r^2  M_{BH}^{2/3}\bigg(1-\frac{2  M_{BH}^{1/3}}{r}\bigg)^2}{a^2}~. 
\end{equation}
Up to $\mathcal{O}(1/a^{2})$ and using the mass profile described above, we get the $g^{rr}$ component of the metric as follows:
\begin{equation}\label{g_lovlock10}
g^{rr}\equiv g(r)= \left(1-\frac{2 M_{BH}^{1/3}}{r}\right) \left(1-\frac{2 M r \left(1-\frac{2 M_{BH}^{1/3}}{r}\right)}{a^2}\right) ~. 
\end{equation}
Similarly, we can also calculate the matter density profile from the mass profile described above (\ref{massprofile_lovlock10}). From the temporal component of the Einstein equation $G^{t}{}_{t}=8 \pi T^{t}{}_{t}$, we get the matter density in the following form,
\begin{eqnarray}
    4 \pi \rho &=& \frac{m'(r)}{r^8}\nonumber\\
    \Rightarrow \rho &=&\frac{3 M M_{BH}^{2/3}\left( r-2 M_{BH}^{1/3}\right)}{2 \pi  a^2 r^8}~.
\end{eqnarray}
The tangential pressure can be derived through the utilization of the Bianchi identities \cite{Chakraborty:2021dmu}, and its specific expression is as follows,
\begin{eqnarray}
    8 P_{T} = \frac{(m(r))^{1/3}\rho}{r- 2 (m(r))^{1/3}}~.
\end{eqnarray}
\begin{figure}
	\centering
	\minipage{0.78\textwidth}
	\includegraphics[width=\linewidth]{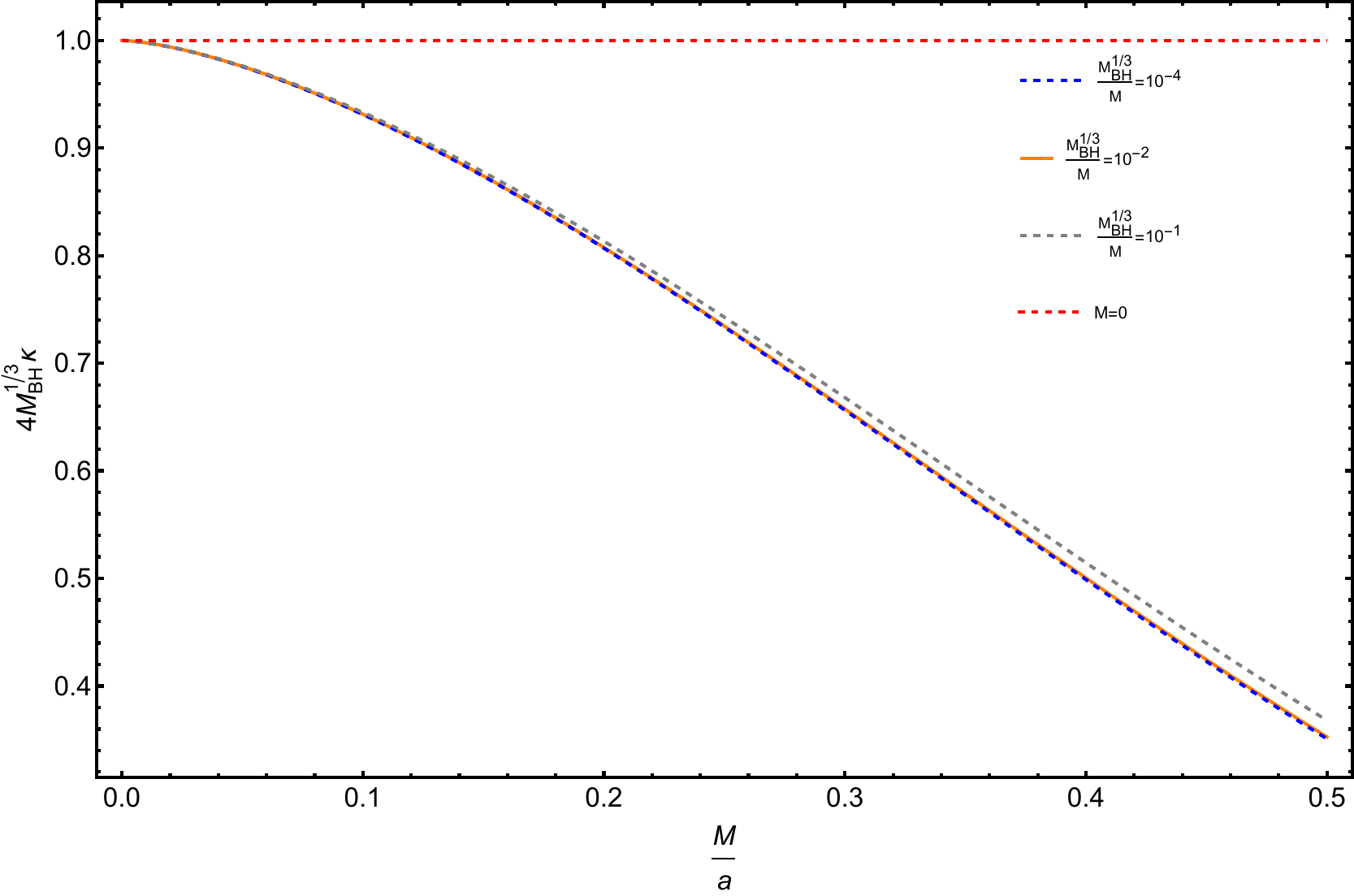}
	\endminipage\hfill
	\caption{Here we have shown the variation of surface gravity with $\frac{M}{a}$ for various galactic parameters.}
	\label{Surface-LOVE-D10}
\end{figure}


One can calculate the radius of the photon sphere, the shadow radius, and the ISCO for a galactic ten-dimensional pure Lovelock black hole of Lovelock order three. Upto the leading order in $\left(\frac{M_{BH}^{1/3}}{a}\right)$, the radius of the photon sphere is given by,
\begin{equation}
   r_{\rm ph}=3 M_{BH}^{1/3}\bigg(1+\frac{M M_{BH}^{1/3}}{a^2}\bigg)~.\label{Lovelockgal_Photon1}  
\end{equation}
One can show that at the limit of $ M\to 0$, one can get back the result of the photon sphere of the isolated ten-dimensional pure Lovelock black hole of Lovelock order three. The above result implies that the presence of a dark matter halo increases the radial location of the photon sphere, compared to the case of an isolated ten-dimensional pure Lovelock black hole of Lovelock order three. The radial coordinate solution of ISCO is given by,
\begin{align}\label{Lovelock-ISCO1}
r_{\rm ISCO}=6M_{BH}^{1/3}\bigg(1-\frac{32 M M_{BH}^{1/3}}{a^2}\bigg)~.
\end{align}
Similarly,  as $M$ approaches zero, we can obtain the result of the ISCO of the isolated ten-dimensional pure Lovelock black hole of Lovelock order three. We show that the presence of a dark matter halo decreases the radial location of the ISCO, compared to the case of an isolated ten-dimensional pure Lovelock black hole of Lovelock order three. The shadow radius associated with the galactic ten-dimensional pure Lovelock blackhole of Lovelock order three is given by,
\begin{align}\label{Lovelock-Shadow1}
r_{\rm sh}=3 \sqrt{3} M_{BH}^{1/3}\bigg[1+\frac{M}{a}+\frac{M(5M-18 M_{BH}^{1/3})}{6 a^2}\bigg]~.
\end{align}
In the scenario where $M$ approaches zero, the outcome reverts to the shadow radius corresponding to the isolated ten-dimensional pure Lovelock black hole of Lovelock order three. Notably, in the presence of a dark matter halo, the shadow radius will be larger than that of an isolated ten-dimensional pure Lovelock black hole of Lovelock order three.

Given our focus on black hole spacetime, the surface gravity becomes well-defined. In the case of static spacetime, its definition is governed by the following relationship,
\begin{align}\label{Killing-Horizon}
    \xi^{\alpha}_{t}\nabla_{\alpha}\xi^{\mu}_{t}=\kappa\xi^{\mu}_{t}~\text{(evaluated on the horizon)}~.
\end{align}
Here $\kappa$ is the surface gravity of the black hole horizon, and $\xi^{\alpha}_{t}$ is the timelike killing vector of the spacetime. For $d=10$ and $N=3$ dimensional galactic pure Lovelock blackhole, this turns out to be,
\begin{align}\label{Surface-Gravity-d10}
  \kappa=e^{\frac{3}{2} \beta }\frac{\sqrt{1+\frac{4MM_{BH}^{1/3}}{a^{2}}}}{4M_{BH}^{1/3}}~.
\end{align}
In the above expression, we have introduced $\beta$ as a shorthand for the quantity, 
\begin{align}\label{def_beta}
\beta=\sqrt{\frac{M}{2a-M}}\left[2\tan^{-1}\left(\frac{a-M}{\sqrt{M(2a-M)}}\right)-\pi\right]~.
\end{align}
The Hawking temperature is defined as $T_{H}=\kappa/2 \pi$. Thus, from reference \ref{Surface-LOVE-D10}, it is observed that the introduction of a dark matter halo causes a reduction in the Hawking temperature when compared to the scenario of an isolated ten-dimensional pure Lovelock black hole of Lovelock order three.

\section{Stability of galactic pure Lovelock black holes under scalar perturbation}\label{sec-4}

In this section, we will study the stability of a galactic pure Lovelock black hole under scalar perturbation. At first, we consider a galactic
seven-dimensional pure Lovelock black hole of Lovelock order two and then a galactic ten-dimensional pure Lovelock black hole of Lovelock order three. Here, we wish to study how a massless scalar field $\Psi$ evolves in the background geometry, with appropriate mass profiles for the dark matter environment in the context of the pure Lovelock blackhole and the galactic pure Lovelock blackhole, respectively. We will assume that the energy density of the scalar field $\Psi$ is small, such that it can be ignored compared to the energy densities of various matter species appearing in the galactic pure Lovelock black hole spacetimes. Thus, the scalar field $\Psi$ can be considered as a test scalar field, living in the background geometry of the galactic pure Lovelock blackhole and satisfying the massless Klein-Gordon equation, $g^{\mu\nu}\nabla_{\mu}\nabla_{\nu}\Psi=0$. For the metric $g_{\mu \nu}$, we can exploit the result that it describes a static and spherically symmetric geometry, such that the scalar field admits the following decomposition,
\begin{align}\label{Lovelock-Scalar-decom}
\Psi(t,r,\Omega)=\frac{1}{r^{(d-2)/2}}\sum_{l=0}^{\infty}\sum_{m=-l}^{l}e^{-i\omega t}Y_{lm}(\Omega)\psi_{lm}(r)~.
\end{align}
Here, $Y_{lm}(\Omega)$ corresponds to the spherical harmonics, and $\psi_{lm}(r)$ is the radial function that needs to be determined. Until now, we have not used any explicit form for the metric describing the background spacetime. Substituting the above decomposition for the scalar field $\Psi$ in the massless Klein-Gordon equation, we obtain the following equation of the radial part of the scalar field,
\begin{align}\label{Lovelock-Master}
\frac{d^{2}\psi_{lm}}{dr_{*}^{2}}+\Big[\omega^{2}-V_{l}(r)\Big]\psi_{lm}=0~.
\end{align} 
Here we have defined the tortoise coordinate $r_{*}$ as the solution of the following differential equation, 
\begin{align}\label{tortoiseLovelock}
\frac{dr_{*}}{dr}=\frac{1}{\sqrt{g^{rr}g_{tt}}}=\sqrt{\frac{1}{f(r)g(r)}}~,
\end{align} 
and the potential $V_{l}(r)$ is given by, 
\begin{align}\label{gen_pot_lovelock}
V_{l}(r)= f (r) \bigg[\frac{l(l+d-3)}{r^2}+\frac{(d-2)(d-4)}{4 r^2}g(r)+\frac{(d-2)}{4 r}\bigg(g^{\prime}(r)+\frac{f^{\prime}(r)}{f(r)}g(r)\bigg)\bigg]~.
\end{align} 
Now, to determine the characteristic quasi-normal mode frequencies of the pure Lovelock black hole and galactic pure Lovelock black hole spacetimes, it is essential to furnish explicit expressions for the mass function and the function $f(r)$. These expressions are contingent on the specific solution under consideration, and they will vary between a galactic seven-dimensional pure Lovelock black hole of Lovelock order two and a ten-dimensional pure Lovelock black hole of Lovelock order three. Consequently, we will independently conduct a stability analysis for these two galactic pure Lovelock black hole spacetimes in the subsequent sections.

\subsection{Stability of the galactic pure Lovelock black hole}\label{sec-4.1}

We begin by discussing the stability of the galactic pure Lovelock black hole spacetime. Initially, we examine the mass profile $m(r)$ (given in \ref{massprofile_lovlock7} ) and the $f(r)$ component (given in \ref{Lovelock7_gal_f}). After substituting both of these expressions into \ref{gen_pot_lovelock}, we get the explicit form of the potential, which is provided in the \ref{app-A}.

To proceed further, we must also determine an explicit expression for the tortoise coordinate $r_{*}$. Since this requires integration, the analysis has a constant of integration. To ensure consistency with a galactic pure black hole horizon at $r=2 \sqrt{M_{BH}}$, where $r_{*}(r=2 \sqrt{M_{BH}})=-\infty$, we carefully choose the integration constant. In this case, we use $r_{*}\in (-\infty,\infty)$ to cover both the horizon and the asymptotic region, where $r\rightarrow\infty$ represents the asymptotic region and $r=2 \sqrt{M_{BH}}$ represents the horizon, respectively. By imposing the condition that $a>M>>M_{BH}>0$, we get,
\begin{align}\label{appx-tor_love10}
r_{*}\simeq \exp\left[\frac{-\beta}{2}\right] \bigg(1- \frac{2 M \sqrt{M_{BH}}}{a^2}\bigg)\left[r+2 \sqrt{M_{BH}} \log \bigg(\frac{r}{2 \sqrt{M_{BH}}}-1\bigg)\right]~.
\end{align}
Then, we consider the mass profile $m(r)$ in  \ref{massprofile_lovlock10}, and the $f(r)$ component by \ref{Lovelock10_gal_f}. Substituting both of these expressions in \ref{gen_pot_lovelock},  we get the explicit form of the potential provided in the \ref{app-A}.
To proceed further, besides the potential, we also need to provide an explicit expression for the tortoise coordinate $r_{*}$. Since determining the tortoise coordinate involves an integration, the analysis has a constant of integration. We choose the integration constant, keeping in mind that there is a galactic pure Lovelock black hole horizon at $r=2 M_{BH}^{1/3}$, such that $r_{*}(r=2 M_{BH}^{1/3})=-\infty$. In this case, we will use $r_{*}\in (-\infty,\infty)$ to cover both the horizon and the asymptotic region, such that $r\rightarrow\infty$ represents the asymptotic region and $r=2 M_{BH}^{1/3}$ represents the horizon. Using the condition that $a>M>>M_{BH}>0$, we obtain,
\begin{align}\label{appx-tor_love}
r_{*}\simeq \exp\left[\frac{-\beta}{2}\right] \bigg(1- \frac{2 M M_{BH}^{1/3}}{a^2}\bigg)\left[r+2 M_{BH}^{1/3} \log \bigg(\frac{r}{2 M_{BH}^{1/3}}-1\bigg)\right]~.
\end{align}

\subsubsection{Ringdown waveform of galactic pure Lovelock black hole}\label{sec-4.1.2}

After providing a comprehensive overview of the perturbation equation related to the ringdown spectrum, we now focus on the methodology used to obtain the quasi-normal mode (QNM) frequencies and, subsequently, the time-domain ringdown signal. This signal will then be compared with that of an isolated pure Lovelock black hole. To determine the QNM frequencies, we employ the 6th-order WKB approximation \cite{Konoplya:2004ip}. Here, we have used the following boundary conditions: ingoing modes at the event horizon and outgoing modes at the asymptotic infinity. The QNM frequencies for both the isolated pure Lovelock black hole and the galactic pure Lovelock black hole can be found in \ref{Galatic_Lovelock_7_Scalar}, \ref{Galatic_Lovelock_7}, \ref{Galatic_Lovelock_10_Scalar} and \ref{Galatic_Lovelock_10}, respectively. We have noted that the imaginary parts of the QNM frequencies for the galactic pure Lovelock black hole are smaller than their isolated counterpart, implying that the pure Lovelock black hole is less stable in the presence of a galaxy.

\begin{table}[th!]
	\centering
	\def\arraystretch{1.3}
	\setlength{\tabcolsep}{1.5em}
	\begin{tabular}{|p{1cm}||p{5cm}|p{5cm}|  }
	\hline
\multicolumn{3}{|c|}{Comparison of QNM frequencies} \\
	\hline
	Mode $ n $    & For isolated pure Lovelock black hole      & For galactic pure Lovelock black hole     \\ \hline
		$ 0  $    & $0.5300 -0.0938 i $           & $0.5250 -0.0919 i$                     \\ \hline
		$ 1  $    & $0.5114 -0.2854 i$            & $0.5069 -0.2797 i$                      \\ \hline
		$ 2  $    & $0.4780 -0.4889 i $           & $0.4745 -0.4788 i$                       \\ \hline
		$ 3  $    & $0.4385 -0.7095 i$            & $0.4350 -0.6932 i$                        \\ \hline
		$ 4  $    & $0.4042 -0.9442 i $           & $0.3963 -0.9224 i$                         \\ \hline
		$ 5  $    & $0.3735 -1.1773 i$           & $0.3639 -1.1625 i$                          \\ \hline
        $ 6  $    & $0.3453-1.4112 i$            & $0.3403 -1.4087 i$                           \\ \hline
		$ 7  $    & $0.3124 -1.6397 i $           & $0.3265 -1.6566 i$                            \\ \hline
		\end{tabular}
	\caption{A comparison between the real and the imaginary parts of the QNM frequencies of the seven-dimensional isolated and the seven-dimensional galactic pure Lovelock black hole of Lovelock order two has been presented for $\ell=1$. We have chosen the galactic parameters such that $\frac{M}{a}=10^{-2}$. In this case, the values are more or less in agreement, with the isolated pure Lovelock black hole being more stable than the galactic one.}\label{Galatic_Lovelock_7_Scalar}
\end{table}


\begin{table}[th!]
	\centering
	\def\arraystretch{1.3}
	\setlength{\tabcolsep}{1.5em}
	\begin{tabular}{|p{1cm}||p{5cm}|p{5cm}|  }
	\hline
\multicolumn{3}{|c|}{Comparison of QNM frequencies} \\
	\hline
	Mode $ n $    & For isolated pure Lovelock black hole      & For galactic pure Lovelock black hole     \\ \hline
		$ 0  $    & $0.7348-0.0949 i $           & $0.7276-0.0930 i$                     \\ \hline
		$ 1  $    & $0.7210-0.2869 i$            & $0.7143-0.2812 i$                      \\ \hline
		$ 2  $    & $0.6953-0.4855 i $           & $0.6892-0.4757 i$                       \\ \hline
		$ 3  $    & $0.6608-0.6944 i$            & $0.6557-0.6799 i$                        \\ \hline
		$ 4  $    & $0.6225-0.9151 i $           & $0.6179-0.8954 i$                         \\ \hline
		$ 5  $    & $0.5859-1.1453 i $           & $0.5804-1.1214 i$                          \\ \hline
            $ 6  $    & $0.5528-1.3819 i$            & $0.5465-1.3557 i$                           \\ \hline
		$ 7  $    & $0.5236-1.6207 i $           & $0.5175-1.5951 i$                            \\ \hline
		\end{tabular}
	\caption{A comparison between the real and the imaginary parts of the QNM frequencies of the seven-dimensional isolated and the seven-dimensional galactic pure Lovelock black hole of Lovelock order two has been presented for $\ell=2$. We have chosen the galactic parameters such that $\frac{M}{a}=10^{-2}$. In this case, the values are more or less in agreement, with the isolated pure Lovelock black hole being more stable than the galactic one.}\label{Galatic_Lovelock_7}
\end{table}


\begin{table}[th!]
	\centering
	\def\arraystretch{1.3}
	\setlength{\tabcolsep}{1.5em}
	\begin{tabular}{|p{1cm}||p{5cm}|p{5cm}|  }
	\hline
\multicolumn{3}{|c|}{Comparison of QNM frequencies} \\
	\hline
	Mode $ n $    & For isolated pure Lovelock black hole      & For galactic pure Lovelock black hole     \\ \hline
		$ 0 $     & $0.7330-0.0912 i$            & $0.7259 -0.0894 i$                     \\ \hline
		$ 1 $     & $0.7189-0.2756 i$            & $0.7121 -0.2701 i$                      \\ \hline
		$ 2 $     & $0.6917-0.4660 i$            & $0.6859 -0.4566 i$                       \\ \hline
		$ 3 $     & $0.6542-0.6662 i$            & $0.6494 -0.6521 i$                        \\ \hline
		$ 4 $     & $0.6104-0.8794 i$            & $0.6066 -0.8590 i$                         \\ \hline
		$ 5 $     & $0.5673-1.1071 i$            & $0.5614-1.0778 i$                          \\ \hline
            $ 6 $     & $0.5282-1.3446 i$            & $0.5175-1.3075 i$                           \\ \hline
		$ 7 $     & $0.5042-1.5916 i$            & $0.4778-1.5460 i$                            \\ \hline
		
	\end{tabular}
	\caption{A comparison between the real and the imaginary parts of the QNM frequencies of the ten-dimensional isolated and the ten-dimensional galactic pure Lovelock black hole of Lovelock order three has been presented for $\ell=1$. We have chosen the galactic parameters such that $\frac{M}{a}=10^{-2}$. In this case, the values are more or less in agreement, with the isolated pure Lovelock black hole being more stable than the galactic one.}\label{Galatic_Lovelock_10_Scalar}
\end{table}


\begin{table}[th!]
	\centering
	\def\arraystretch{1.3}
	\setlength{\tabcolsep}{1.5em}
	\begin{tabular}{|p{1cm}||p{5cm}|p{5cm}|  }
	\hline
\multicolumn{3}{|c|}{Comparison of QNM frequencies} \\
	\hline
	Mode $ n $    & For isolated pure Lovelock black hole      & For galactic pure Lovelock black hole     \\ \hline
		$ 0 $     & $0.9513-0.0929 i$            & $0.9419-0.0911 i$                     \\ \hline
		$ 1 $     & $0.9404-0.2800 i$            & $0.9314-0.2744 i$                      \\ \hline
		$ 2 $     & $0.9195-0.4709 i$            & $0.9110-0.4615 i$                       \\ \hline
		$ 3 $     & $0.8897-0.6680 i$            & $0.8821-0.6544 i$                        \\ \hline
		$ 4 $     & $0.8532-0.8732 i$            & $0.8465-0.8552 i$                         \\ \hline
		$ 5 $     & $0.8129-1.0875 i$            & $0.8069-1.0647 i$                          \\ \hline
            $ 6 $     & $0.7714-1.3108 i$            & $0.7659-1.2832 i$                           \\ \hline
		$ 7 $     & $0.7312-1.5420 i$            & $0.7259-1.5097 i$                            \\ \hline
		
	\end{tabular}
	\caption{A comparison between the real and the imaginary parts of the QNM frequencies of the ten-dimensional isolated and the ten-dimensional galactic pure Lovelock black hole of Lovelock order three has been presented for $\ell=2$. We have chosen the galactic parameters such that $\frac{M}{a}=10^{-2}$. In this case, the values are more or less in agreement, with the isolated pure Lovelock black hole being more stable than the galactic one.}\label{Galatic_Lovelock_10}
\end{table}

\begin{figure}
	\centering
	\minipage{0.68\textwidth}
	\includegraphics[width=\linewidth]{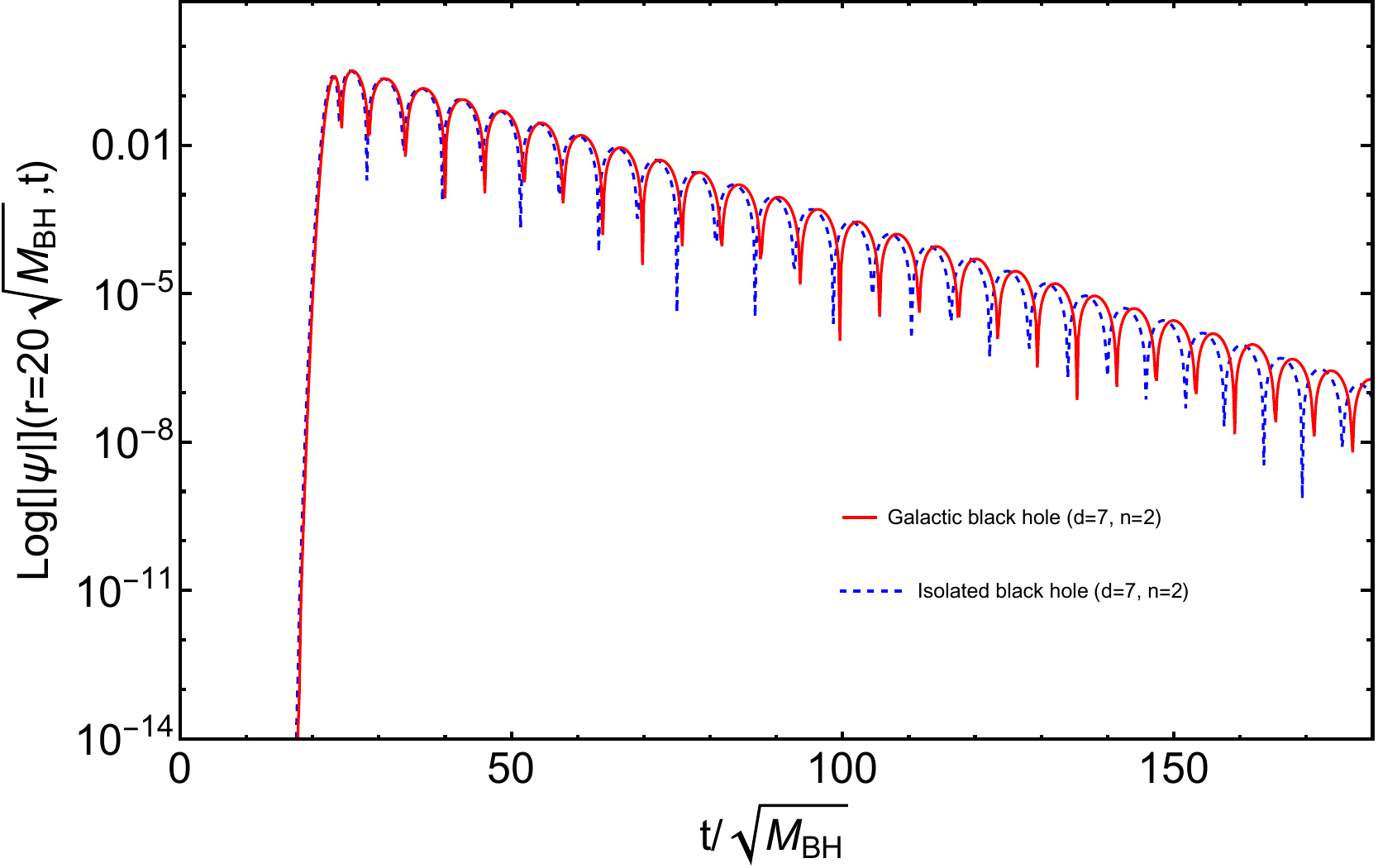}
	\endminipage\hfill
	\minipage{0.68\textwidth}
	\includegraphics[width=\linewidth]{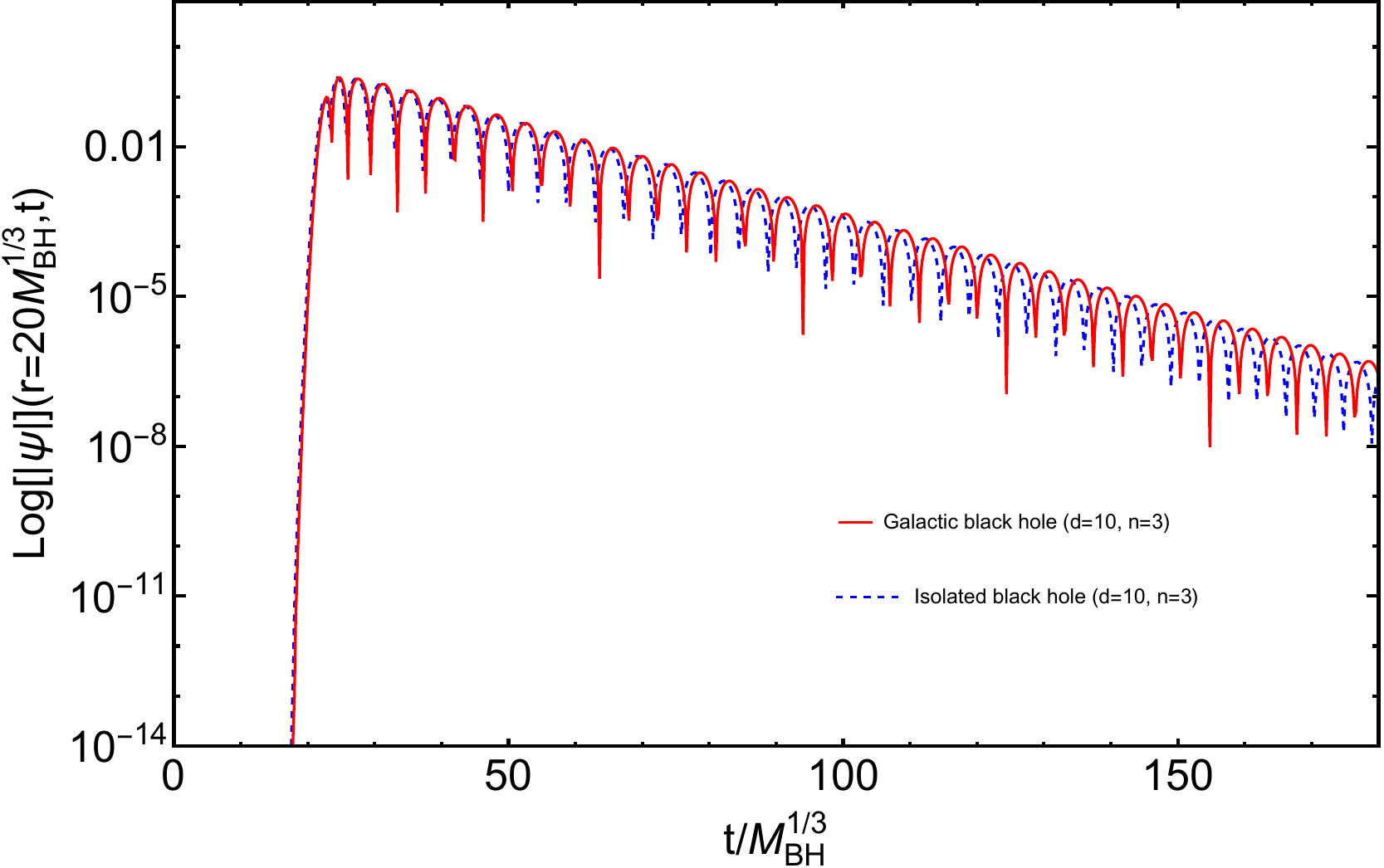}
	\endminipage
	\caption{This figure is dedicated to a comparison of the ringdown waveform of galactic pure Lovelock black holes to its isolated counterpart. In the left figure, we have considered $d=7$ and $N=2$. Similarly, we have considered $d=10$ and $N=3$ in the right figure. For both the plots, we have used \ref{Initial conditions1}, such that $r^{0}_{*}$ lies outside of the corresponding photon sphere. For both cases, we have chosen the galactic parameters such that $\frac{M}{a}=10^{-2}$.} 
	\label{LOVE-LOG}
\end{figure}

In order to derive the time-domain signal for the ringdown profile of the galactic pure Lovelock black hole, our focus shifts to analyzing the evolution of the scalar perturbation in the time domain. To accomplish this, we define the Fourier counterpart of $\psi_{lm}(r_{*},\omega)$ in the usual manner. Utilizing the perturbation in the time domain and taking into account that the effective potential in \ref{Lovelock-Master} remains frequency-independent for a static and spherically symmetric spacetime, we can reexpress the radial perturbation equation from \ref{Lovelock-Master} as follows:
\begin{align}\label{Time-domain-EQ1}
-\frac{\partial^2\psi_{lm}(r_{*},t)}{\partial t^2}+\frac{\partial^{2}\psi_{lm}(r_{*},t)}{\partial r_{*}^{2}}-V_{l}(r_{*})\psi_{lm}(r_{*},t)=0~.
\end{align}
The given equation is a second-order partial differential equation concerning both $r_{*}$ and $t$, and solving it requires defining two initial conditions in time and two boundary conditions in the tortoise coordinate. For the boundary conditions, outgoing conditions are imposed at asymptotic infinity, while ingoing conditions are applied at the horizon. These boundary conditions can be expressed as $\partial_{r_*}\psi_{lm}(r_{*},t)= \mp \partial_{t}\psi_{lm}(r_{*},t)$ as $r_*\to \pm \infty$. As for the initial conditions, we choose the following:
\begin{equation}\label{Initial conditions1}
\psi_{lm}(r_*,0)=0\quad\text{and}\quad\partial_{t}\psi_{lm}(r_{*},0)=e^{-\frac{(r_{*}-r^{0}_{*})^{2}}{\sigma^{2}}}~.
\end{equation}
We carefully select $r^{0}_{*}$ and $\sigma^{2}$ to ensure that the primary signal manifests outside the unstable photon sphere, explicitly avoiding the maxima of the potential. With this appropriately chosen boundary and initial conditions, we proceed to solve \ref{Time-domain-EQ1} numerically. The outcomes of this analysis are available in \ref{LOVE-LOG}, where we present the ringdown waveform in the time domain. This time domain single can allow us to identify the galactic
parameters, as well as to distinguish them from the GR counterparts if it is observed in future generations of gravitational wave measurements.

\section{Discussion}\label{sec-5}

In this paper, we first-time model galactic black holes within the framework of pure Lovelock gravity. The spacetime metric is obtained by considering the Hernquist-type mass profile for the galaxy, resulting in a pure Lovelock black hole situated at the galaxy's center. The focus is on the $d=3N+1$ galactic pure Lovelock black hole of Lovelock order $N$. The study reveals that the presence of the galaxy affects the photon sphere and the shadow radius, causing an increase compared to the isolated counterpart. In contrast, the radial location of the ISCO decreases compared to the isolated counterpart. The Hawking temperature for the same setup has been computed. It has been demonstrated that the presence of a dark matter halo decreases the Hawking temperature compared to the scenario of an isolated pure Lovelock black hole.

The impact of the dark matter halo on the geometry of these galactic pure Lovelock black holes is also investigated, and it turns out to significantly influence their stability. The presence of galactic matter leads to a modification of the potential experienced by the perturbing scalar field, which affects the prompt ringdown. The study shows that the imaginary parts of the QNM frequencies for the galactic pure Lovelock black hole are smaller compared to those of the isolated pure Lovelock black hole, indicating that the presence of a galaxy renders the pure Lovelock black hole less stable. This information could be valuable in identifying galactic parameters as well as distinguishing them from their isolated pure Lovelock counterparts and the GR counterparts theoretically or in future experimental observations but the effect of extra dimension in pure Lovelock gravity must be reduced to $d = 4$ by some mechanism (e.g., compactification). Our future work aims to extend this analysis to a general $d$ dimension of the Lovelock black hole of Lovelock order $N$.

While this work utilizes the Hernquist-type mass profile to describe the galactic pure Lovelock black hole spacetimes, a more comprehensive mass profile, such as the NFW (Navarro–Frenk–White) profile \cite{Navarro:1995iw}, could also be employed. However, obtaining closed-form analytic solutions with the NFW mass profile might be challenging. Nevertheless, it could offer a more realistic depiction of the impact of the dark matter halo on pure Lovelock black hole geometries, and further exploration of this matter is intended for future research.

\section*{Acknowledgments}
 CS thanks the Saha Institute of Nuclear Physics (SINP) Kolkata for financial support. SB thanks IACS for financial support. CS is thankful to
IACS, for their warm hospitality and research facilities.

\appendix
\labelformat{section}{Appendix #1} 
\labelformat{subsection}{Appendix #1}
\section{The complete mass profile of a \texorpdfstring{$d=3 N+1$}{Lg} dimensional galactic pure Lovelock black hole of Lovelock order \texorpdfstring{$N$}{Lg} }\label{app-1}
We get the complete mass profile for a seven-dimensional pure Lovelock black hole of Lovelock order two in the galactic center by considering \ref{pressure} and \ref{Lovelock7_gal_f} as,
\begin{align}\label{mass_lovelock7}
m(r)&=\frac{A}{B}~,
\end{align}
where $A= M_{BH} \bigg( a^2 \sqrt{2 a-M+4 M_{BH}}+\frac{M^{2} r \bigg(2-\frac{r}{\sqrt{M_{BH}}}\bigg)}{\sqrt{2 a-M+4 M_{BH}}}+r^2 \sqrt{2 a-M+4 M_{BH}}+ \frac{4 M M_{BH} r \bigg(\frac{r}{\sqrt{M_{BH}}}-2\bigg)}{\sqrt{2 a-M+4 M_{BH}}}\\+2 a r \Bigg( \frac{M \bigg(\frac{r}{\sqrt{M_{BH}}}-2\bigg)}{\sqrt{2 a-M+4 M_{BH}}}+\sqrt{2 a-M+4 M_{BH}}\Bigg)+2 M (2 M_{BH}-r) \sqrt{2 a-M+4 M_{BH}}\bigg)^2$, and \\ $B= \bigg(a^2 \sqrt{2 a-M+4 M_{BH}}+\frac{2 M^{2} r \bigg( \frac{2 \sqrt{ M_{BH}}}{r}-1\bigg)}{\sqrt{2 a-M+4 M_{BH}}}+r^2 \sqrt{2 a-M+4 M_{BH}}+\frac{8 M M_{BH} r \bigg(1-\frac{2 \sqrt{ M_{BH}}}{r}\bigg)}{\sqrt{2 a-M+4 M_{BH}}}\\+2 a r \Bigg( \frac{2 M \bigg(1-\frac{2 \sqrt{ M_{BH}}}{r}\bigg)}{\sqrt{2 a-M+4 M_{BH}}}+\sqrt{2 a-M+4 M_{BH}}\Bigg)+2 M (2 M_{BH}-r) \sqrt{2 a-M+4 M_{BH}}\bigg)^2$. Up to $\mathcal{O}(1/a^{2})$, we get the mass profile of the seven-dimensional pure Lovelock black hole of Lovelock order two as in \ref{massprofile_lovlock7}.\\

Similarly, we get the complete mass profile for a ten-dimensional pure Lovelock black hole of Lovelock order three in the galactic center by considering \ref{pressure} and \ref{Lovelock10_gal_f} as,
\begin{align}\label{mass_lovelock10}
m(r)&=\frac{C}{D}~,
\end{align}
where $C=M_{BH} \bigg(a^2 \sqrt{2 a-M+4 M_{BH}}+ \frac{M^{2} r \bigg(2-\frac{r}{M_{BH}^{1/3}}\bigg)}{\sqrt{2 a-M+4 M_{BH}}}+2 a r \Bigg( \frac{M \bigg(\frac{r}{M_{BH}^{1/3}}-2\bigg)}{\sqrt{2 a-M+4 M_{BH}}}+\sqrt{2 a-M+4 M_{BH}}\Bigg)\\+ \frac{4 M M_{BH} r \bigg(\frac{r}{M_{BH}^{1/3}}-2\bigg)}{\sqrt{2 a-M+4 M_{BH}}}+r^2 \sqrt{2 a-M+4 M_{BH}}+2 M (2 M_{BH}-r) \sqrt{2 a-M+4 M_{BH}}\bigg)^3$, and \\ $D=\bigg(a^2 \sqrt{2 a-M+4 M_{BH}}+\frac{2 M^{2} r \bigg( \frac{2 M_{BH}^{1/3}}{r}-1\bigg)}{\sqrt{2 a-M+4 M_{BH}}}+2 a r \Bigg(\frac{2 M \bigg(1-\frac{2 M_{BH}^{1/3}}{r}\bigg) }{\sqrt{2 a-M+4 M_{BH}}}+\sqrt{2 a-M+4 M_{BH}}\Bigg)\\+ \frac{8 M M_{BH} r \bigg(1-\frac{2 M_{BH}^{1/3}}{r}\bigg)}{\sqrt{2 a-M+4 M_{BH}}}+r^2 \sqrt{2 a-M+4 M_{BH}}+2 M (2 M_{BH}-r) \sqrt{2 a-M+4 M_{BH}}\bigg)^3$. Up to $\mathcal{O}(1/a^{2})$, we obtain the mass profile of the ten-dimensional pure Lovelock black hole of Lovelock order three as in \ref{massprofile_lovlock10}.

Also, if we consider the $f(r)$ as in \ref{Lovelock7_gal_f} and mass profile $m(r)$ as in \ref{massprofile_lovlock7}, we get the radial pressure of a seven-dimensional pure Lovelock black hole of Lovelock order two in the galactic center as,
\begin{align}\label{radial_pressure_7}
8 \pi P_{r}&= \frac{2 \sqrt{M_{BH}+\frac{2 M \sqrt{M_{BH}} r^2} {a^2}\bigg(1-\frac{2 \sqrt{ M_{BH}}}{r}\bigg)^2} \Bigg(\frac{2 M \bigg(1-\frac{2 \sqrt{ M_{BH}}}{r}\bigg)}{a^2+2 a r+4 M M_{BH}-2 M r+r^2}+\frac{2 \sqrt{ M_{BH}}}{r^2}\bigg)}{r^4 \bigg(1-\frac{2 \sqrt{ M_{BH}}}{r}\Bigg)} \times \nonumber \\ & \Bigg(1-\frac{2\sqrt{M_{BH}+\frac{2 M \sqrt{M_{BH}} r^2}{a^2} \bigg(1-\frac{2 \sqrt{ M_{BH}}}{r}\bigg)^2}}{r}\Bigg)-\frac{4 \Bigg(M_{BH}+\frac{2 M \sqrt{M_{BH}} r^2}{a^2} \bigg(1-\frac{2 \sqrt{ M_{BH}}}{r}\bigg)^2\Bigg)}{r^6}~.
\end{align}
Up to $\mathcal{O}(1/a^{2})$, we get the radial pressure ($P_{r}$) is zero. Therefore, we can consider the $f(r)$ as in \ref{Lovelock10_gal_f} and mass profile $m(r)$ as in \ref{massprofile_lovlock10} to proceed further.

Also, if we consider the $f(r)$ as in \ref{Lovelock10_gal_f} and mass profile $m(r)$ as in \ref{massprofile_lovlock10}, we get the radial pressure of a ten-dimensional pure Lovelock black hole of Lovelock order three in the galactic center as,
\begin{align}\label{radial_pressure_10}
8 \pi P_{r}&=\frac{3 \bigg(M_{BH}+\frac{3 M M_{BH}^{2/3} r^2}{a^2} \bigg(1-\frac{2 M_{BH}^{1/3}}{r}\bigg)^2\bigg)^{2/3} \bigg(\frac{2 M  \bigg(1-\frac{2 M_{BH}^{1/3}}{r}\bigg)}{a^2+2 a r+4 M M_{BH}-2 M r+r^2}+\frac{2 M_{BH}^{1/3}}{r^2}\bigg)}{r^7 \bigg(1-\frac{2 M_{BH}^{1/3}}{r}\bigg)}\times\nonumber\\ & \bigg(1-\frac{2 \bigg(M_{BH}+\frac{3 M M_{BH}^{2/3} r^2}{a^2} \bigg(1-\frac{2 M_{BH}^{1/3}}{r}\bigg)^2\bigg)^{1/3}}{r}\bigg)-\frac{6 \bigg(M_{BH}+\frac{3 M M_{BH}^{2/3} r^2}{a^2} \bigg(1-\frac{2 M_{BH}^{1/3}}{r}\bigg)^2\bigg)}{r^9}~.
\end{align}
Up to $\mathcal{O}(1/a^{2})$, we get the radial pressure ($P_{r}$) is zero. Therefore, we can consider the $f(r)$ as in \ref{Lovelock10_gal_f} and mass profile $m(r)$ as in \ref{massprofile_lovlock10} to proceed further.

\section{Efeective potential for a galactic pure Lovelock black hole}\label{app-A} 

The complete expression of the effective potential experienced by a test scalar field living in a seven-dimensional galactic pure Lovelock black hole of Lovelock order two is given by,
\begin{align}\label{Effective_potetial_lovelock7}
V_{l}(r)& =e^{\gamma}\Bigg(1-\frac{2 \sqrt{M_{BH}}}{r}\Bigg) \Bigg(\frac{l (l+4)}{r^2}+\frac{5\sqrt{M_{BH}+\frac{2 M \sqrt{M_{BH}}} {a^2}\Bigg(r-2 \sqrt{M_{BH}}\Bigg)^2}}{2 r^3}\nonumber\\ &+\frac{15 \Bigg(r-2 \sqrt{M_{BH}+\frac{2 M \sqrt{M_{BH}}} {a^2}\Bigg(r-2 \sqrt{M_{BH}}\Bigg)^2}\Bigg)}{4 r^3}+\frac{5 M \Bigg(2 M_{BH}-\sqrt{M_{BH}} r\Bigg)}{a^2 r^2 \sqrt{M_{BH}+\frac{2 M \sqrt{M_{BH}}} {a^2}\Bigg(r-2 \sqrt{M_{BH}}\Bigg)^2}} \nonumber\\ &+\frac{5 \Bigg(M r \Bigg(2 \sqrt{M_{BH}}-r\Bigg)-\sqrt{M_{BH}} \Bigg(a^2+2 a r+4 M M_{BH}-2 M r+r^2\Bigg)\Bigg)}{2 r^3 \Bigg(2 \sqrt{M_{BH}}-r\Bigg) \Bigg(a^2+2 a r+4 M M_{BH}-2 M r+r^2\Bigg)} \times \nonumber\\ &\Bigg(r-2 \sqrt{M_{BH}+\frac{2 M \sqrt{M_{BH}}} {a^2}\Bigg(r-2 \sqrt{M_{BH}}\Bigg)^2}\Bigg)\Bigg)~.
\end{align}
The complete expression of the effective potential experienced by a test scalar field living in a ten-dimensional galactic pure Lovelock black hole of Lovelock order three is given by,
\begin{align}\label{Effective_potetial_lovelock10}
V_{l}(r)& =\frac{e^{\gamma}}{r^2}\bigg(1-\frac{2 M_{BH}^{1/3}}{r}\bigg) \bigg(4 \bigg(\frac{M \bigg(r-2 \sqrt{M_{BH}+\frac{3 M M_{BH}^{2/3}} {a^2} \bigg(r-2 M_{BH}^{1/3}\bigg)^2}\bigg)}{a^2+2 a r+4 M M_{BH}-2 M r+r^2}\nonumber\\ & +M_{BH}^{1/3} \bigg(\frac{2 \sqrt{M_{BH}+\frac{3 M M_{BH}^{2/3}} {a^2} \bigg(r-2 M_{BH}^{1/3}\bigg)^2}}{r \bigg(2 M_{BH}^{1/3}-r\bigg)}+\frac{6 M \sqrt{M_{BH}+\frac{3 M M_{BH}^{2/3}} {a^2} \bigg(r-2 M_{BH}^{1/3}\bigg)^2}}{a^2 M_{BH}^{1/3}+3 M \bigg(r-2 M_{BH}^{1/3}\bigg)^2}+\frac{1}{r-2 M_{BH}^{1/3}}\bigg)\nonumber\\ & -\frac{5 \sqrt{M_{BH}+\frac{3 M M_{BH}^{2/3}} {a^2} \bigg(r-2 M_{BH}^{1/3}\bigg)^2}}{r}-\frac{3 M r \sqrt{M_{BH}+\frac{3 M M_{BH}^{2/3}} {a^2} \bigg(r-2 M_{BH}^{1/3}\bigg)^2}}{a^2 M_{BH}^{1/3}+3 M \bigg(r-2 M_{BH}^{1/3}\bigg)^2}+3\bigg)+l^2+7 l\bigg)~.
\end{align}
Both expressions related to the effective potential experienced by a scalar field have been used in the main text.


\bibliography{Blackhol}

\bibliographystyle{./utphys1}


\end{document}